\documentclass[aps,pre,showpacs,groupedaddress]{revtex4}
\usepackage{amsfonts}
\usepackage{mathrsfs}
\usepackage{graphicx}
\usepackage{amsmath}
\usepackage{amssymb}
\usepackage{amsbsy}

\newcommand{\ignore}[1]{\index{ignore}}

\def\@{\partial}
\def\<{\langle}
\def\>{\rangle}
\def\nn{\nonumber}
\def\bm{\boldsymbol}

\def\eg{{\it e.g.}}
\def\ie{{\it i.e.}}

\def\hh{\hat{h}}
\def\hq{\hat{q}}
\def\ii{\mathrm{i}}
\def\ti{_{it}}
\def\tj{_{jt}}
\def\si{_{is}}
\def\sj{_{js}}

\def\ts{_{ts}}
\def\bh{\boldsymbol{h}}

\def\bs{\bm \sigma}
\def\bq{\boldsymbol{q}}

\def\mL0{\mathcal{L}_0}
\def\vp{\varphi}
\def\sgn{\mbox{sgn}}

\begin{document}

\title{
Structure of attractors in randomly connected networks}
\author{Taro Toyoizumi}
\email{taro.toyoizumi@brain.riken.jp}
\affiliation{RIKEN Brain Science Institute, Wako-shi, Saitama
351-0198, Japan}
\affiliation{Deptartment of Computational Intelligence and Systems Science, Tokyo Institute of Technology, Yokohama 226-8502, Japan}
\author{Haiping Huang}
\email{physhuang@gmail.com}
\affiliation{RIKEN Brain Science Institute, Wako-shi, Saitama
351-0198, Japan}
\date{\today}
\begin{abstract}
The deterministic dynamics of randomly connected neural networks are
studied, where a state of binary neurons evolves according to a
discrete-time synchronous update rule. We give a theoretical support
that the overlap of systems' states between the current and a
previous time develops in time according to a Markovian stochastic
process in large networks. This Markovian process predicts how often
a network revisits one of previously visited states, depending on
the system size. The state concentration probability, \ie, the
probability that two distinct states co-evolve to the same state, is
utilized to analytically derive various characteristics that
quantify attractors' structure. The analytical predictions about the
total number of attractors, the typical cycle length, and the number
of states belonging to all attractive cycles match well with
numerical simulations for relatively large system sizes.
\end{abstract}
\pacs{02.50.Ey, 84.35.+i, 05.45.-a}
 \maketitle
\section{Introduction}
Neurons in the brain interact with each other in a heterogeneous and
asymmetric way~\cite{Nature-11}, producing complex neuronal dynamics
for information processing. In the past decades, there are a surge
of research interests in randomly connected neural
networks~\cite{Rajan-2010,Sompolinsky-1988,Vreeswijk-1996,Taro-2011,Ostojic-2014,Aljadeff-2014}.
Although their behavior is described by simple deterministic
equations, the resulting dynamics are rich, exhibiting fixed-point
behavior, limit cycles, or high-dimensional chaos. These networks
are capable of generating useful dynamic activity patterns after
appropriate learning~\cite{Sussillo-2009,Laje-2013}.

Simple models of neural
networks~\cite{Parisi-1986,Derrida-1987,Young-1988,Rieger-1989,Berg-1992,Crisanti-1993,Bastolla-1997,Amari-2013,Huang-2014}
have been explored to elucidate characteristics of their complex
dynamics. In these networks, connections between binary neurons are
independently drawn from an identical distribution, and the state of
a network is updated simultaneously in discrete time steps without
thermal noise. Thus, every initial configuration must evolve into an
attractor, which is either a fixed point or a limit cycle.  Because
a fixed point is a limit cycle of length $l=1$, the whole state
space is divided into separated basins of attractions with
heterogenous cycle lengths. Extensive numerical simulations were
carried out to analyze the typical cycle length and the number of
cycles~\cite{Young-1988,Nutzel-1991}. The typical length of the
cycles was observed to grow exponentially with the number of neurons
$n$ (such kinds of cycles are called chaotic attractors), and the
total number of attractors increases linearly with $n$. These
quantities were also analytically evaluated based on an empirical
assumption that the dynamics loses memory of its non-immediate
past~\cite{Bastolla-1997}.

In this work, we develop a dynamic mean-field theory to characterize
the attractors of the asymmetric neural network by extending the
state concentration concept~\cite{Amari-2013}, recently introduced
to characterize the robustness and quickness of network's transient
dynamics. Our analysis estimates the (cumulative) distribution for
the cycle length of attractors, the total number of attractors, and
the volume of attractors in the state space.

 We remark that our work has
three-fold contributions for understanding the statistical
properties of the dynamics of randomly connected neural networks.
First, a theoretical support for the Markovian property of state
concentration dynamics (termed the annealed approximation in
Ref.~\cite{Bastolla-1997}) is provided by computing the finite-size
effect of the mean-field theory by explicitly evaluating the
quenched randomness of network connections. Second, we provide a
detailed picture about how state concentration happens in randomly
connected neural networks. In particular, we quantify what is the
characteristic distance that typically leads to state concentration
and evaluate characteristic time scales underlying the state
concentration dynamics. Finally, our theory gives a good consistency
with numerical simulations on the distribution of the cycle length,
the typical cycle length, the number of cycles, and the total number
of states belonging to all attractive cycles. These three
contributions complement the previous
studies~\cite{Young-1988,Nutzel-1991,Berg-1992,Bastolla-1997,Amari-2013}
and provide deep insights towards the dynamics of randomly-connected
neural networks.

The paper is organized as follows. In Sec.~\ref{Model}, we define
the neural network model and its dynamics. Mean-field analysis is
presented in detail in Sec.~\ref{Theory}. Results on the state
concentration and statistical properties of attractors are discussed
in Sec.~\ref{Stcon} and Sec.~\ref{Att}, respectively. We summarize
our results in Sec.~\ref{Conc}.

\section{Model definition}
\label{Model}
 We consider randomly connected neural networks consisting of $n$
 neurons (units). Each unit interacts with all the other units with an
 asymmetric coupling. We use $J_{ij}$ to represent the coupling
 strength from unit $j$ to $i$, and $J_{ij}$ is independent
 of $J_{ji}$ (and others), and they follow the same Gaussian
 distribution with zero mean and variance $1/n$. The state of neuron
 $i\, (i=1,\dots,n)$ at time $t+1\, (t=0,1,\dots)$
 is set according to the parallel deterministic dynamics in
 discrete time steps by its input $h_i(t)$ as
 \begin{equation}\label{phi0}
\sigma_i(t+1)=\sgn (h_i(t))=\begin{cases} +1, &\text{(active state)}\\
-1, &\text{(silent state)}
\end{cases}
\end{equation}
where the input is defined by
 \begin{equation}\label{paradyn}
h_i(t) = \sum_{j=1}^n J_{ij}\sigma_j(t).
\end{equation}
Therefore, by combining Eqs. (\ref{phi0}) and (\ref{paradyn}), the dynamics are summarized by
$\sigma_i(t+1)=\sgn{\left(\sum_{j}J_{ij}\sigma_{j}(t)\right)}$ in
terms of the activity,  or equivalently by
$h_i(t+1)=\sum_{j}J_{ij}\sgn(h_{j}(t))$ in terms of the input.

We later compare the dynamics of randomly connected neural networks
to that of random Boolean networks
\cite{Kauffman-1969a,Kauffman-1969b}, where each one of $2^n$ states
$\bs(t)=\{\sigma_i(t)|i=1,\dots,n\}$ is randomly mapped to another.

\section{Mean-field analysis}
\label{Theory} We study the dynamical evolution of the overlap
between two states along a trajectory, expecting that its
distribution across different realizations of $\{J_{ij}\}$ contains information about the structure of attractors.
Let us define the overlap of two states,
$\bs(t)=\{\sigma_i(t)|i=1,\dots,n\}$ and
$\bs(s)=\{\sigma_i(s)|i=1,\dots,n\}$ along the same trajectory at different times $t>s$, by
\begin{eqnarray}
  \label{eq:q}
  q_{ts}\equiv\frac{1}{n}\sum_{i=1}^n \sigma_i(t)\sigma_i(s).
\end{eqnarray}
This overlap takes $+1$ if two states are the same and $-1$ if one
is the sign-flip of the other. The overlap takes a discrete value for a finite size network,
but can be approximated as a continuous quantity in the large network size limit.
The mean-field theory provides the
dynamics of this overlap parameter and its fluctuation defined over the ensemble of random $\{J_{ij}\}$ (see Appendix~\ref{DynF}). The stochastic dynamics of
the overlap is well approximated for large $n$ by a Markovian
process
\begin{eqnarray}
  \label{eq:markov}
  P_{t+1,s+1}(q)\approx\int  W(q|q')P\ts(q') dq',
\end{eqnarray}
where $P\ts(q)\equiv \mbox{Prob}(q\ts=q)$ is the probability of
$q\ts=q$. The transition probability is approximated for large but
finite $n$ by a simple binomial distribution
\begin{eqnarray}
  \label{eq:W}
  W(q|q')&=&\binom{n}{n(1+q)/2}\left[\frac{1+\varphi(q')}{2}\right]^{\frac{n(1+q)}{2}}\left[\frac{1-\varphi(q')}{2}\right]^{\frac{n(1-q)}{2}}\nn\\
&\approx&\exp\left[n\left(H(q)+\frac{1+q}{2}\ln\frac{1+\vp(q')}{2}+\frac{1-q}{2}\ln\frac{1-\vp(q')}{2}\right)\right],
\end{eqnarray}
where $\vp(q)\equiv (2/\pi)\arcsin q$ and $H(q)\equiv
-\frac{1+q}{2}\ln\frac{1+q}{2}-\frac{1-q}{2}\ln\frac{1-q}{2}$.  Note
that Eq.~(\ref{eq:W}) summarizes the probability that $n(1+q)/2$ out
of $n$ neurons take the same sign in state $\bs(t+1)$ and
$\bs(s+1)$, given that $n(1+q')/2$ out of $n$ neurons take the same
sign in the previous step. The binomial distribution in
Eq.~(\ref{eq:W}) suggests that the state overlap for each neuron is
approximately independent, occurring with probability
$(1+\varphi(q'))/2$ (see Appendix~\ref{DynF} for a support).

A similar expression is obtained for random Boolean networks by
replacing $\vp(q)$ with $\vp_{BN}(q)\equiv\delta_{q,1}$, simply
reflecting completely random nature of state transitions.

It is worth noting that, the dynamics of the overlap becomes
deterministic in the limit of large $n$ according to the central limit theorem, which is the so called distance
law~\cite{Derrida-1986c,Derrida-1986d,Derrida-1987b,Amari-1974,Kurten-1988},
$q_{t+1,s+1}=\vp(q\ts)$. In this equation, the equality
holds only at $q=0$ and $q=\pm 1$, and otherwise $|\vp(q)| < |q|$. Hence, in the limit of large $n$, the
overlap monotonically converges to the stable solution of $q=0$,
implying that two distinct states would never converge. On the other
hand, for finite $n$, the overlap fluctuates with amplitude $\sim
1/\sqrt{n}$ about the deterministic solution (see detailed
explanations in Appendix~\ref{DynF}). Thus, the overlap can evolve
from $q<1$ to $q=1$ in time, indicating that system's state
eventually comes back to one of previously visited states in a
finite network.

The Markovian process of Eq.~(\ref{eq:markov}) sequentially provides
$P_{t+l,t}(q)$ for $t=1,2,\dots$ for some positive time difference
$l=t-s$ given an initial distribution at $t=0$. The initial distribution is
denoted by $P_{l,0}(q)\equiv {\rm Prob}(q_{l,0}=q)$. Since the
initial state, $\bs(0)$, is selected randomly and independently from
$\{J_{ij}\}$, we can set without losing generality the initial state
to be $\sigma_i(0)=1$ for all $i$ (see Appendix~\ref{INIT}). In this
case, the initial overlap of interest is expressed by
\begin{eqnarray}
\label{eq:ql0}
q_{l,0}&=&\frac{1}{n}\sum_{i=1}^n \sigma_i(l) \sigma_i(0)\nn \\
&=&\frac{1}{n}\sum_{i=1}^n \sgn\left(h_i(l-1)\right).
\end{eqnarray}
If $l$ is small, $P_{l,0}(q)$ reflects the memory of the initial
state $\bs(0)$ and is hard to evaluate exactly.  However, if $l$ is
large, the mean-field result in Appendix~\ref{DynF} indicates that
$\{h_i(l-1)| i=1,2,\dots,n\}$ follows approximately a zero-centered
independent Gaussian distribution with unit variance in the large
network-size limit. This means that the state overlap of
Eq.~(\ref{eq:ql0}) approaches a distribution centered around zero
with variance $\sim 1/n$. In particular, $P_{l,0}(q)$ tends for
large $l$ to a binomial distribution
$\binom{n}{n(1+q_{l,0})/2}2^{-n}$, where the probability of
$q_{l,0}=\pm 1$ is approximately $2^{-n}$ in the large network-size
limit. We confirm this property later with numerical simulations.

\section{State concentration}
\label{Stcon} In this section, we consider how different states
concentrate in time. The Markovian dynamics of Eq.~(\ref{eq:markov})
are completely characterized by the eigenvalues and eigenvectors of
the transition probability $W$ \cite{Gal-2014}. Let
$f_a(q)$ and $\lambda_a \, (\le 1)$ respectively be the $a$th
eigenvector and eigenvalue of $W$.  We rank eigenvalues in a
descending order, \ie, $\lambda_1\ge\lambda_2\ge\dots\ge\lambda_{n+1}$ (the number of possible values for $q$ is $n+1$). The
distribution of the overlap is expressed by a weighted sum of the
eigenvectors as
\begin{eqnarray}
  P_{t+l,t}(q) = \sum_{a=1}^{n+1} (\lambda_a)^t A_a f_a(q),
\end{eqnarray}
where $\{A_a\}$ is a set of initial coefficients that satisfies
$P_{l,0}(q)=\sum_{a}A_a f_a(q)$. Hence, as the time step increases,
$P_{t+l,t}(q)$ becomes progressively dominated by the components with
large eigenvalues.

\begin{figure}[h!]
\includegraphics[scale=1.0]{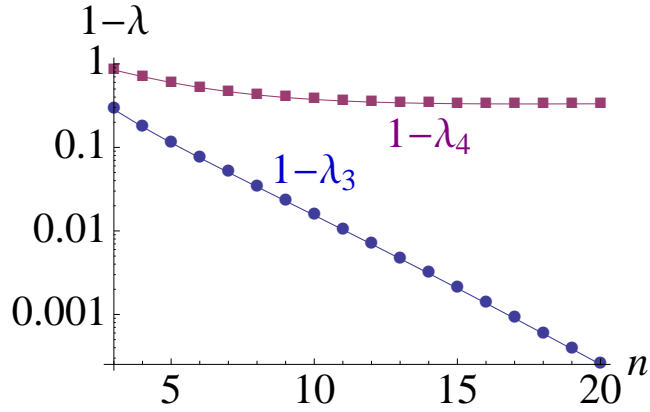}
\caption{ (Color online) The non-trivial eigenvalues $\lambda_3$ and
$\lambda_4$ of the transition matrix $W$.
  }\label{fig:lam}
\end{figure}

It is easy to see that $W$ has two trivial eigenvectors
$f_1(q)=\delta_{q,1}$ and $f_2(q)=\delta_{q,-1}$ with degenerate
eigenvalues $\lambda_1=\lambda_2=1$. Note that $\delta_{q,q_0}$ is
the Kronecker delta function. The third eigenvector $f_3(q)$ is a
non-trivial one and its eigenvalue $\lambda_3\approx 1-\exp(-0.41n)$
exponentially approaches 1 with $n$ (see Fig.~\ref{fig:lam} for the
numerical result). The fourth eigenvalue converges to
$\lambda_4\approx0.67$ in the limit of large $n$. The half-decay
time of the $a$-th component is described by these eigenvalues and
given by $t_a\equiv (\ln 2)/(-\ln \lambda_a)$, or equivalently,
$(\lambda_a)^{t_a}=1/2$. There is a clear gap between the decay time
of the third and fourth eigen-components. This result indicates
that, for large $n$, the distribution of the overlap must approach
quickly a quasi-stationary state $P_*(q)\equiv\sum_{a=1}^3 A_a
f_a(q)$ at around $t_4\approx 1.73$ and stay unchanged until
$t_3\approx 0.69\exp(0.41n)$. In particular, the quasi-stationary
state is characterized solely by $f_3(q)$ except at $q=\pm 1$.

 This
analysis also suggests when the mean-field theory breaks down ---
the theory is not applicable once the third eigen-component
significantly decays at around $t_3$. That is, after an exponential
time of $\exp(0.41n)$, the distribution of the overlap becomes the
linear combination of $f_1(q)$ and $f_2(q)$, \ie, every state
becomes either the same or the sign-flip of the others. However,
this never happens in a real system.

In the remaining part of this
section, we characterize in more detail the quasi-stationary state in large $n$ limit,
from which we extract the structure of attractors.

 We first introduce an
auxiliary notation
\begin{eqnarray}
  \label{eq:alpha}
  \alpha_{t+l,t}(q)\equiv \frac{1}{n}\ln P_{t+l,t}(q),
\end{eqnarray}
where $\int \exp(n\alpha_{t+l,t}(q))dq=1$ according to the
normalization constraint. With this notation, we can express the
dynamics of Eq.~(\ref{eq:markov}) by
\begin{eqnarray}
\label{eq:dalpha}
  \alpha_{t+l+1,t+1}(q)&=&\frac{1}{n}\ln \int W(q|q')P_{t+l,t}(q')dq'\nn\\
&\approx&
H(q)+\max_{q'}\left[\frac{1+q}{2}\ln\frac{1+\vp(q')}{2}+\frac{1-q}{2}\ln\frac{1-\vp(q')}{2}+\alpha_{t+l,t}(q')\right],
\end{eqnarray}
where the Laplace's method was applied in the second line assuming
large $n$. Note that, in the above expression, the maximizer $q'$ of
the second term is a function of $q$. In particular, the
well-defined asymptotic solution of Eq.~(\ref{eq:dalpha}), \ie,
\begin{eqnarray}
\label{eq:alphass}
  \alpha_*(q) =  H(q)+\max_{q'}\left[\frac{1+q}{2}\ln\frac{1+\vp(q')}{2}+\frac{1-q}{2}\ln\frac{1-\vp(q')}{2}+\alpha_*(q')\right],
\end{eqnarray}
with finite $\alpha_*(q)$ self-consistently provides the
quasi-stationary state. Note that Eq. (\ref{eq:alphass}) permits
arbitrary discontinuity of $\alpha_*(q)$ at $q=\pm 1$, reflecting
that $q=\pm 1$ is the sink of the Markovian process. However, in the
following analysis, we assume continuous $\alpha_*(q)$.

Next, we define index $\beta_{t+l,t}(q\ts)\equiv \frac{1}{n}\ln
\mbox{Prob}(q_{t+l,t}|q_{t+l+1,t+1}=1)$ that characterizes the
probability that two states $\bs(t+l)$ and $\bs(t)$ have overlap
$q_{t+l,t}$ before converging in the next step ($q_{t+l+1,t+1}=1$).
This index is expressed, using the Bayes theorem, in terms of
$\alpha$ by
\begin{eqnarray}
  \label{eq:beta}
  \beta_{t+l,t}(q')&=&\frac{1}{n}\ln \frac{W(1|q')P_{t+l,t}(q')}{P_{t+l+1,t+1}(1)}\nn\\
  &=&\ln\frac{1+\vp(q')}{2}+\alpha_{t+l,t}(q')-\alpha_{t+l+1,t+1}(1).
\end{eqnarray}
This means that, for large $n$, most of the trajectories that lead
to state concentration had an overlap specified by the peak location
of $\beta_{t+l,t}$, \ie, $\arg\max_{q'}\beta_{t+l,t}(q')$, in the
previous step.

\begin{figure}[h!]
\begin{tabular}{ll}
{\bf (a)} & {\bf (b)}\\
\includegraphics[width=8cm]{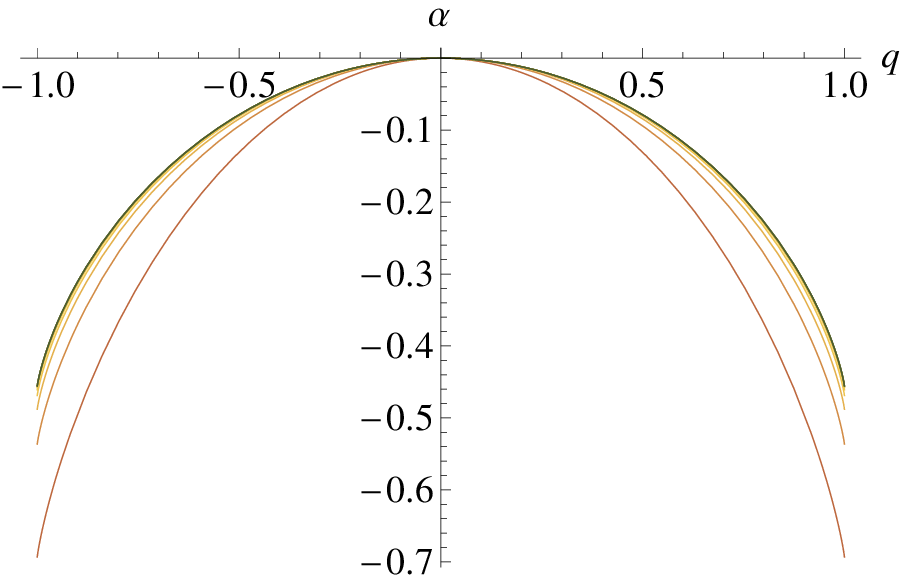}&
\includegraphics[width=8cm]{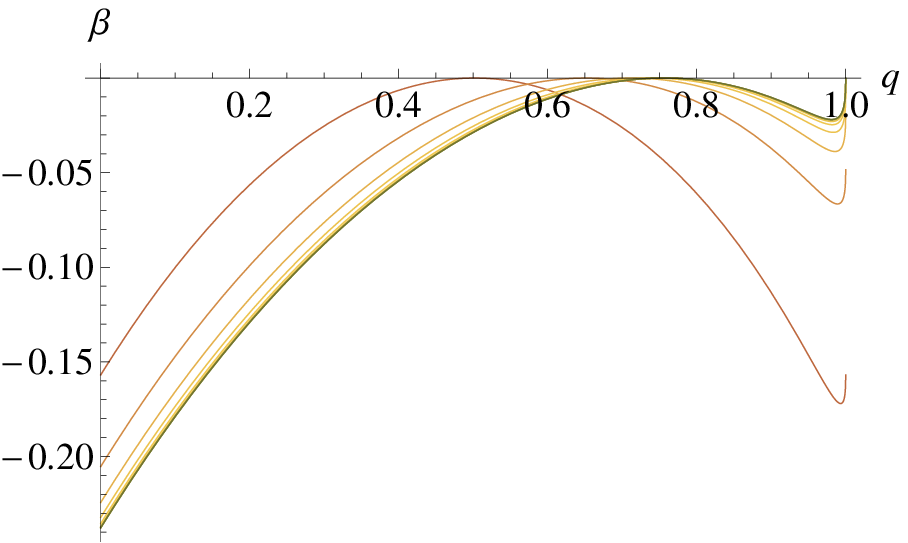}
\end{tabular}
\caption{(Color online) The Markovian dynamics of
  $\alpha_{t+l,t}$ and $\beta_{t+l,t}$ in time.  (a) The index $\alpha_{t+l,t}$
  characterizes the dynamics of distribution $P(q_{t+l,t})$. The line
  color changes from the lowest curve (the orange curve at $t=0$, i.e., $\alpha_{l,0}$ or $\beta_{l,0}$)
  to the yellow, and finally to the gray (the top curve at $t=10$, i.e., $\alpha_{l+10,10}$ or $\beta_{l+10,10}$). (b)
  The index $\beta_{t+l,l}$ characterizes the dynamics of distribution
  $P(q_{t+l,t}|q_{t+l+1,t+1}=1)$. The result indicates that
  states concentrate mainly from $q\approx 0.5$ at the beginning
  but concentrate equally from $q\approx 0.75$ and $q=1$ at the quasi-stationary state.
  We used $\alpha_{l,0}(q)=H(q)-\ln 2$ as the initial condition assuming no correlations at starting points.
  The results hold for any $l\ge1$.
  }\label{fig:alpha_beta}
\end{figure}

In the case of randomly connected neural networks studied here,
$\beta_{t+l,t}(q)$ has two peaks. As shown in Fig.~\ref{fig:alpha_beta} (b), one peak is located at $q=1$ reflecting 
the monotonic increase in $\ln \frac{1+\vp(q)}{2}$ toward $q=1$ and the other peak is 
located at $q<1$ reflecting the peak of $\alpha_{t+l,t}(q)$ at $q=0$ in Eq.~(\ref{eq:beta}). 
The $q<1$ peak shifts to a larger positive value of $q$ and its amplitude loses the dominance over the $q=1$ peak 
as $t$ increases because $\alpha_{t+l,t}(q)$ becomes blunt at large $t$ (Fig.~\ref{fig:alpha_beta} (a)). 
The two peaks become comparable at around $t_4$. In finite-size
systems, the two peaks become indistinguishable once the difference
of the peak values becomes less than $1/n$. The result indicates
that states concentrate mainly from $q\approx 0.5$ at the beginning
but concentrate equally from $q\approx 0.75$ and $q=1$ at the
quasi-stationary state.

These dynamics of the state overlap reflects the specific structure of
attractors as we shall show below. In contrast to the above situation, for
trivial dynamical systems that converge to a unique fixed-point (\eg,
$h_i(t+1)=(1+h_i(t))/2$), $\beta_{t+l,t}(q)$ has a unique peak, which
tends to approach $q=1$ at large $t$, indicating that most
states concentrate from nearby locations. On the other
hand, in random Boolean networks, states concentrate randomly
from any overlap values. Because most states are orthogonal to each
other for large $n$, states mainly concentrate from $q\approx
0$ (see Appendix~\ref{RBN}).

\section{Statistical properties of attractors}
\label{Att} In this section, we analytically describe the
statistical properties of attractors for randomly connected neural
networks using the state concentration
probability~\cite{Amari-2013}. The state concentration probability
$p_{t+1,s+1}$ that characterizes the conditional probability of
$\bs(t+1)=\bs(s+1)$
 given that no states up to time $t$ along the trajectory are the same or the sign-flip of the
others. Because of the symmetry, $p_{t+1,s+1}$ also characterizes
the probability of $\bs(t+1)=-\bs(s+1)$ given the same condition.
Hence,
\begin{eqnarray}
  p_{t+1,s+1} &\equiv& \mbox{Prob}(q_{t+1,s+1}= \pm 1| \{q_{t',s'}\ne \pm 1 | t'\le t, s'<t'\}).
\end{eqnarray}
This state concentration probability is further approximated under
the Markovian approximation of Eq.(\ref{eq:markov}) by
\begin{eqnarray}
  p_{t+1,s+1} &\approx& \int_{q\ts\ne \pm1}  W(q_{t+1,s+1}=1|q\ts) P(q\ts) dq\ts \nn\\
&=& \exp(n \alpha_{t+1,s+1}(1)), \label{eq:scp}
\end{eqnarray}
which directly follows from Eq.~(\ref{eq:dalpha}). Note that, based
on the consideration of the previous section, we used in the second
line that the result is not sensitive to the exclusion of $q'=\pm 1$
from the integral for large $n$. This is because the $\max_{q'}$ in
Eq.~(\ref{eq:dalpha}) is insensitive to its argument at $q'=\pm 1$
unless the initial distribution $P_{l,0}(q)$ is sharply peaked at
$q=\pm 1$, which is not the case here ({\it c.f.}
Eq.~(\ref{eq:ql0})).

Hence, based on the Markovian property, the probability that the
dynamics starting from $\bs(0)$ comes back for the first time to
$\bs(0)$ after $l$ steps without visiting any sign-flip of
previously visited states is described for large $n$ by
\begin{eqnarray}
  \tilde{P}(l) &\equiv& \mbox{Prob}\left(\{q_{1,0}\ne \pm 1\}, \{q_{2,s}\ne \pm 1|s=0,1\},\cdots, \{q_{l-1,s}\ne \pm 1|s=0,1,\dots,l-2\}, \, q_{l,0}=1\right)\nn\\
&=& (1-2 p_{1,0})\prod_{s=0}^1(1-2 p_{2,s})\cdots\prod_{s=0}^{l-2}(1-2p_{l-1,s}) p_{l,0} \nn\\
&=& p_{l,0}\exp\left(\sum_{t=1}^{l-1}\sum_{s=0}^{t-1}\ln(1-2
p_{t,s}) \right). \label{eq:cyc_length}
\end{eqnarray}
Note that, in the second line of Eq.~(\ref{eq:cyc_length}), the
factor $\prod_{s=0}^{t-1}(1-2 p_{t,s})$ describes the probability
that the state makes a transition at time $t$ to a state distinct
from $\{\pm \bs(s)|s=0,1,\dots,t-1\}$. The final factor, $p_{l,0}$,
describes the probability of coming back to the initial state
$\bs(0)$ after $l$ steps.

\begin{figure}[h!]
\includegraphics[width=8cm]{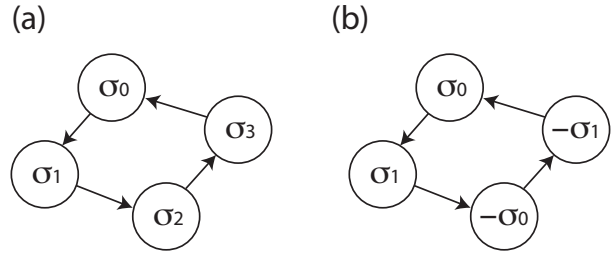}
\caption{There are two kinds of limit cycles if the cycle length $l$
  is even. (a) In the first kind of cycles, the cycle closes without
  ever visiting the sign-flip of previously visited states.  (b) In
  the second kind of cycles, the state first makes a transition to the
  sign-flip of the initial state after $l/2$ steps, \ie,
  $\bs(l/2)=-\bs(0)$. If this happens, the cycle must close after $l$ steps.
  }\label{fig:FigAttractor}
\end{figure}

Altogether, the probability that a certain state, $\bs(0)$, belongs
to a cycle of length $l$ (revisiting $\bs(0)$ for the first time
after $l$ steps) is described for large $n$ by \cite{Bastolla-1997}
\begin{eqnarray}
  P(l) &=& \left\{\begin{array}{cc}
    \tilde{P}(l), & (\mbox{odd $l$})\\
    \tilde{P}(l) + \tilde{P}(l/2).& (\mbox{even $l$})
    \end{array}\right.
\label{eq:cyc_length02}
\end{eqnarray}
Notably, the probability takes different expressions for odd and
even $l$. If $l$ is odd, Eq.~(\ref{eq:cyc_length}) directly gives
the probability. If $l$ is even, there are two separate kinds of
contributions depicted in Fig.~\ref{fig:FigAttractor}. The first
contribution is from cycles that close without ever visiting the
sign-flip of their history. The second contribution is from cycles
that involve a transition at step $l/2$ to the sign-flip of their
initial state, which then guarantees that the cycle closes in $l$
steps.

\begin{figure}[h!]
\includegraphics[bb=2 2 533 413, scale=0.45]{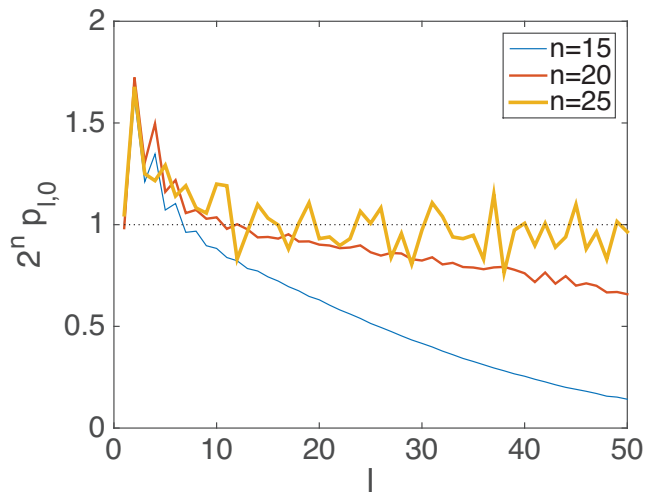}
\caption{(Color online) Simulation results of the initial state
concentration probability $p_{l,0}$
 as a function of $l$ when $n$ varies.
 The results are obtained based on statistics collected
 from $2\times10^9$ networks. The initial state is always set to
 $\sigma_i(0)=1$ for all $i$ (see Appendix~\ref{INIT}).
 The sampling error increases with $n$ because of exponential increase
 of the state space. Note that $p_{1,0}=2^{-n}$ is an exact result.
  }\label{qL0}
\end{figure}

The final step is to evaluate the state concentration probability
$p_{t,s}$. The initial state concentration probabilities are simply
given by
\begin{eqnarray}
p_{l,0}\approx 2^{-n}\equiv p_{\rm init}
\label{eq:pinit}
\end{eqnarray}
for large $n$ and $l$ as discussed in Sec.~\ref{Theory}. Although
this approximation is inaccurate for $l<10$, it becomes accurate for
large $n$ over a wide range of $l$ that includes the typical cycle
length (Fig.~\ref{qL0}).

On the other hand, the
state concentration probability $p_{t+l,t}$ at $t\ge 1$ is computed
sequentially by Eqs.~(\ref{eq:dalpha}) and (\ref{eq:scp}). In
particular, this probability quickly converges within several steps
($t_c\approx 5$; see, Fig.~\ref{fig:alpha_beta} (a)) to the
quasi-stationary value of
\begin{eqnarray}
\label{eq:p}
  p_{\infty}&\equiv& \lim_{t\to\infty}p_{t+l,t} = \exp\left(n \alpha(1)\right)
\end{eqnarray}
for any $l\ge 1$, where $\alpha(1)=-0.46$ from Eq.~(\ref{eq:alphass}). That is, the state concentration
probability quickly converges in several steps from the initial value of $p_{\rm init}\approx\exp(-0.69n)$
to the asymptotic value $p_{\infty}\approx \exp(-0.46n)$.

Therefore, $\tilde{P}(l)$ of Eq.~(\ref{eq:cyc_length}) can be
further approximated using $p_{\rm init}$ and $p_{\infty}$ by
\begin{eqnarray}
  \tilde{P}(l) &=&  p_{\rm init}\exp\left[\sum_{t=1}^{l-1}\sum_{s=0}^{t-1}\ln(1-2p_{\infty}) + O\left(2t_c l\frac{p_{\infty}-p_{\rm init}}{1-2p_{\infty}}\right)\right]\nn\\
&\approx&  p_{\rm init}\exp\left[\frac{l^2}{2}\ln(1-2p_{\infty}) \right]\nn\\
&=&  p_{\rm init}\exp\left(-\frac{l^2}{\tau^2}\right),
\label{eq:cyc_length2}
\end{eqnarray}
where $\tau\equiv \sqrt{-2/\ln(1-2p_\infty)}$ is the characteristic
cycle length that grows exponentially with the system size,
consistent with the numerical observations~\cite{Nutzel-1991}. Note that, in the first line of
Eq.~(\ref{eq:cyc_length2}), we used the relationship that
$|p_{\infty}-p_{ts}|\le |p_{\infty}-p_{\rm init}|$ (for any $t$ and
$s$; see, Fig.~\ref{fig:alpha_beta}) to upper-bound the deviation of
$p_{ts}$ from $p_{\infty}$. To make the contribution of the
 $O\left(2t_c l\frac{p_{\infty}-p_{\rm init}}{1-2p_{\infty}}\right)$ negligible,
 the approximation in the second line assumes
\begin{eqnarray}
4 t_c \frac{p_{\infty}-p_{\rm init}}{-(1-2p_\infty)\ln(1-2p_\infty)}
\ll l \;\; {\rm and} \;\; l\ll
\frac{1-2p_\infty}{2t_c(p_\infty-p_{\rm init})}. \label{eq:cond}
\end{eqnarray}
 The first
condition in Eq.~(\ref{eq:cond}) requires that
$-\frac{l^2}{2}\ln(1-2p_{\infty})\gg2t_c l\frac{p_{\infty}-p_{\rm
init}}{1-2p_{\infty}}$, while the second condition ensures that
$2t_c l\frac{p_{\infty}-p_{\rm init}}{1-2p_{\infty}}\ll1$. The range
of $l$ specified by Eq.~(\ref{eq:cond}) is roughly $10 \ll l\ll
\exp(0.46n)/(2t_c)$
 at $n> 10$.
Hence, the characteristic cycle length $\tau\approx\exp(0.23 n)$ is
well within this range.
Incidentally, $\tau$ is known to also characterize the typical
transient time scale to enter a limit cycle~\cite{Bastolla-1997}.

\begin{figure}[h!]
\begin{tabular}{ll}
{\bf (a)} & {\bf (b)}\\
 \includegraphics[bb=58 24 743 541,width=8cm]{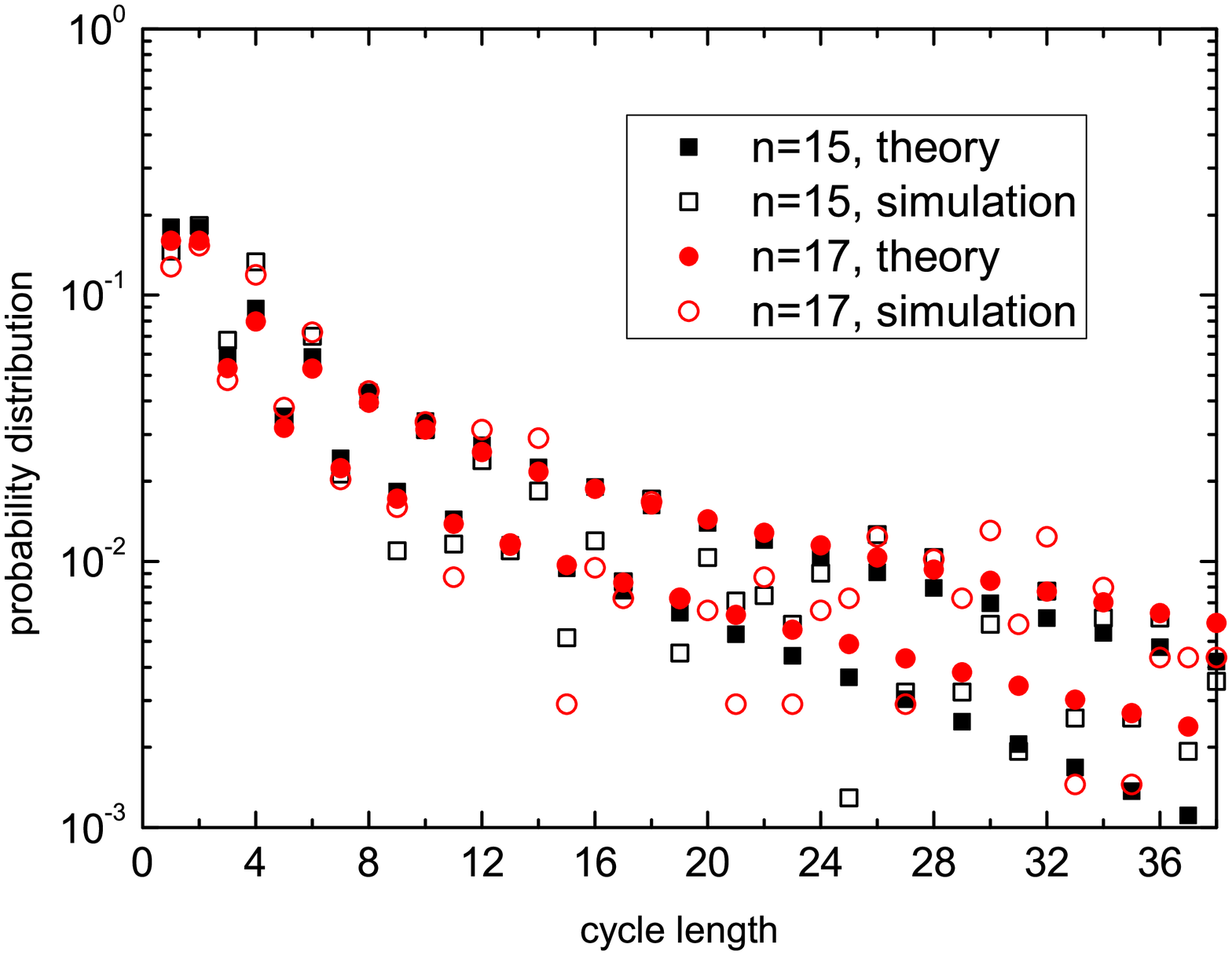}&
\includegraphics[bb=65 26 735 541,width=8cm]{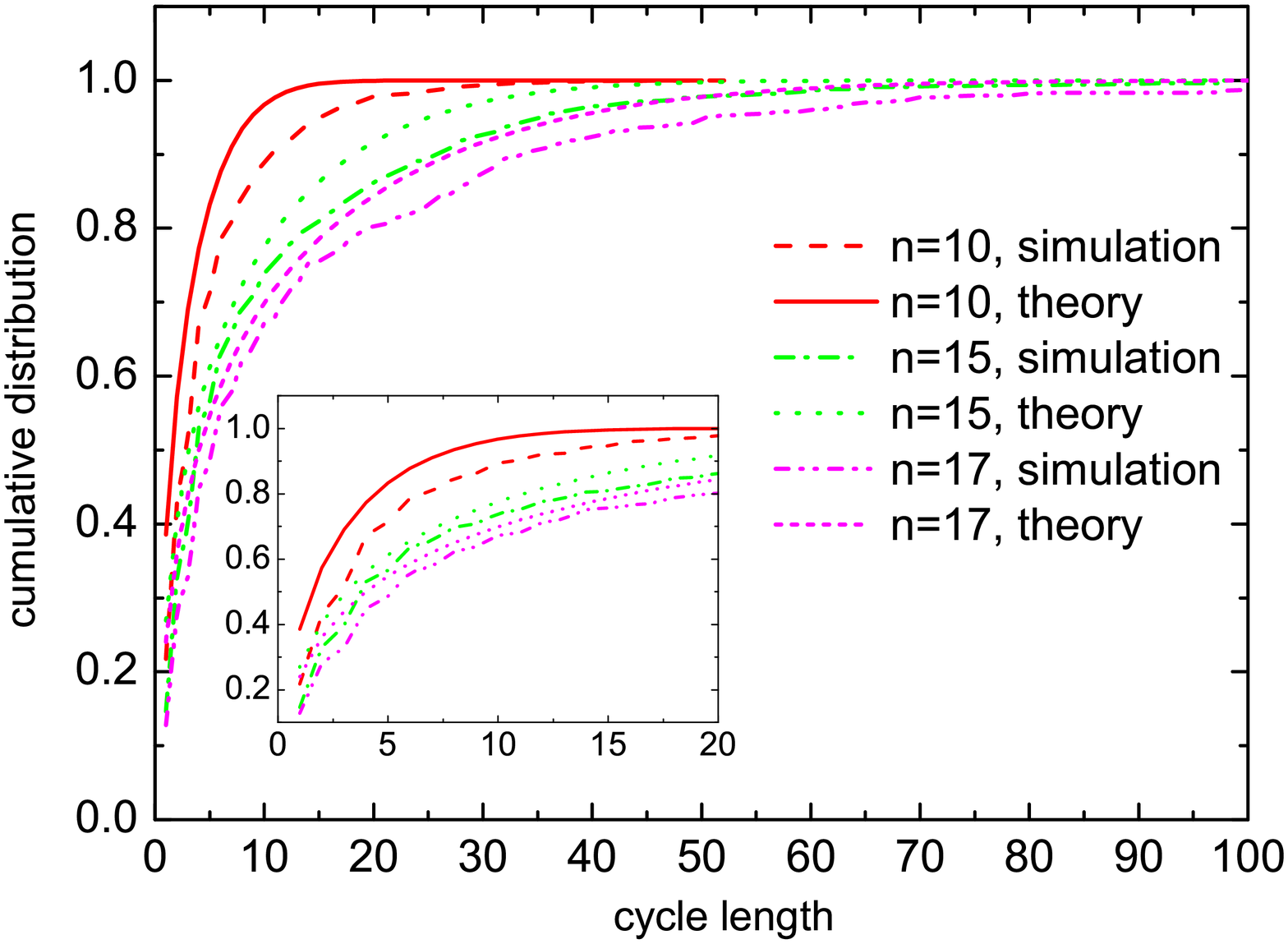}
\end{tabular}
\caption{(Color online) (a) Probability distribution of cycle
lengths. (b) Cumulative distribution of cycle lengths. The numerical
data is obtained from $1000$ samples for $n=10$, $500$ samples for
$n=15$, and $200$ samples for $n=17$. The inset shows an enlarged
view at small cycle length.
  }\label{fig:cum}
\end{figure}

\begin{figure}[h!]
\begin{tabular}{ll}
{\bf (a)} & {\bf (b)}\\
 \includegraphics[bb=78 24 735
543,width=8cm]{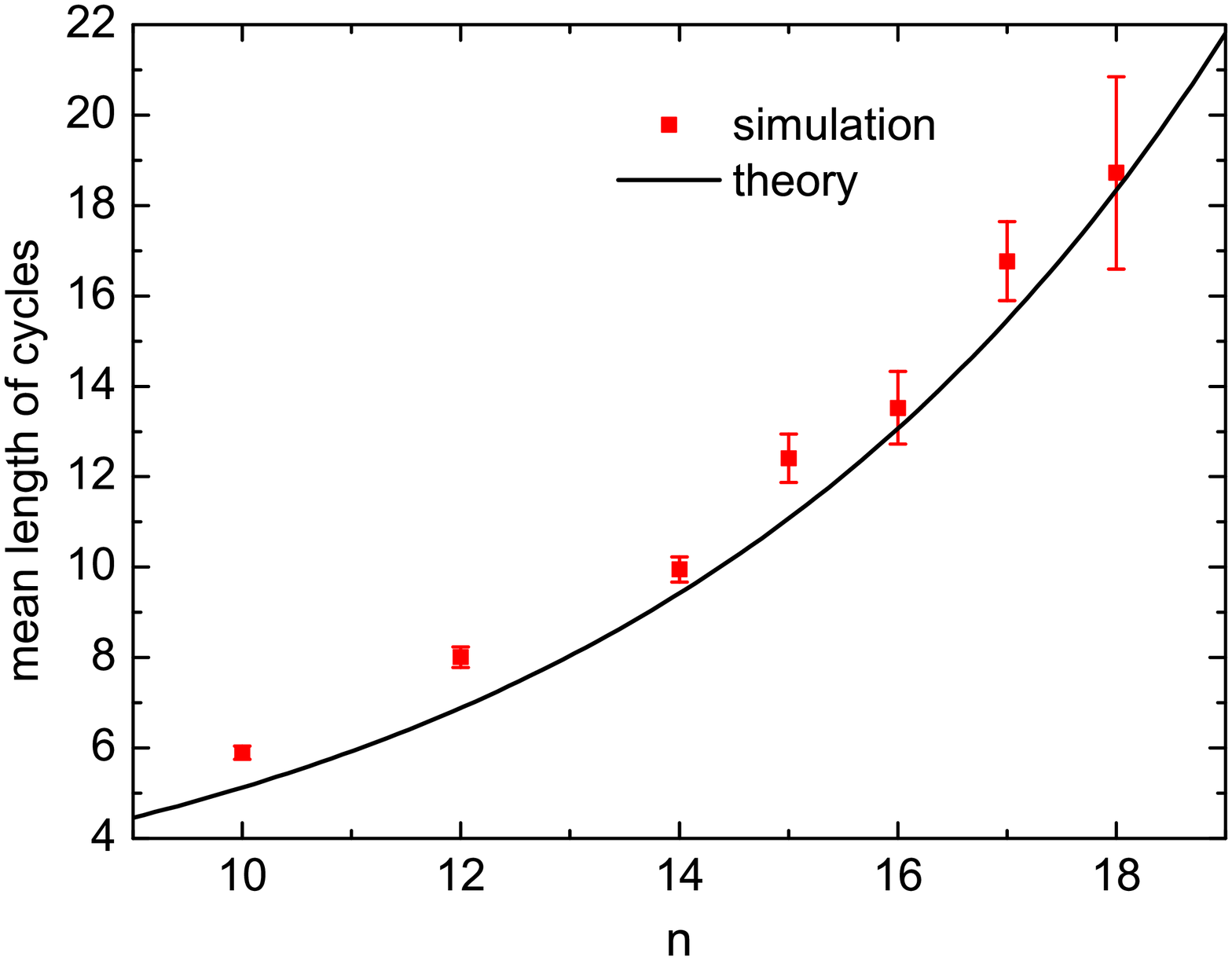}&
\includegraphics[bb=49 23 737 542,width=8cm]{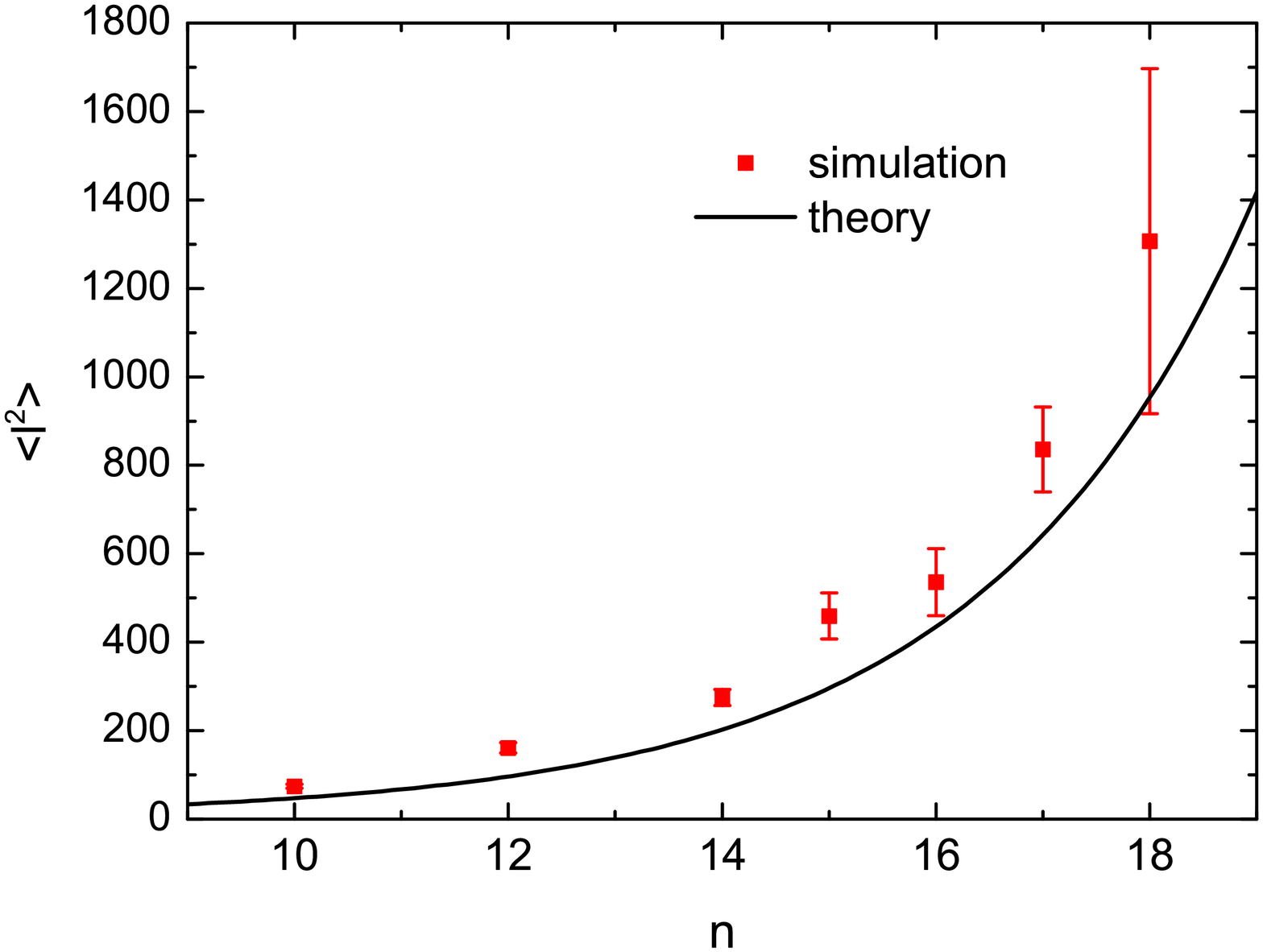}
\end{tabular} \caption{(Color online) The first (mean) and second moment of
the cycle length distribution. Theoretical predictions and numerical
simulations are compared. The results are averaged over many random
realizations of the networks (from $1000$ samples for $n=10$ to
$100$ samples for $n=18$).}\label{fig:cycle}
\end{figure}

The probability of observing a cycle of length $l$ is given by
$P(l)/(l Z)$ with a normalization constant $Z \equiv
\sum_{l=1}^{2^n} P(l)/l$. In this expression, the probability,
$P(l)$, of a state belonging to a cycle of length $l$ should be
divided by $l$ to provide the cycle length probability since all
states within a cycle share the same cycle length.  Note that the
normalization constant $Z$ represents the probability of a state
belonging to a cycle (attractor). Figure~\ref{fig:cum} (a) shows the
comparison of numerically obtained cycle length probability with its
theoretical estimate.  Numerical details to collect the statistics
of the attractors are given in the Appendix~\ref{NumS}.  The theory
nicely captures this probability at around the characteristic cycle
length, including the difference in probability for odd and even
cycle lengths, as $n$ becomes large. However, the deviation is large
for non-typical $l$ in finite networks. The cumulative distribution
of cycle length is similarly obtained by $F(l)\equiv
(1/Z)\sum_{l'=1}^l P(l')/l'\approx
\left(\int_{1}^{l}\tilde{P}(l')/l' {\rm d}l' +
\int_1^{l/2}\tilde{P}(l')/(2l') {\rm d}l'\right)/Z$ . The comparison
of $F(l)$ with the numerical results is shown in Fig.~\ref{fig:cum}
(b). The discrepancy tends to become small for larger $n$ (see the
inset of Fig.~\ref{fig:cum} (b)).

The first moment (mean value) and the second moment of the
distribution can be computed analytically as well. Their values are
given by:
\begin{eqnarray}\label{moments}
  \left<l\right> &=& \frac{4\sqrt{\pi}\tau\left[1-{\rm erf}(1/\tau)\right]}{3\int^{\infty}_{1/\tau^{2}}\frac{e^{-t}}{t}{\rm d}t}, \\
   \left<l^{2}\right> &=& \frac{2\tau^{2}e^{-1/\tau^{2}}}{\int^{\infty}_{1/\tau^{2}}\frac{e^{-t}}{t}{\rm
   d}t},
\end{eqnarray}
where ${\rm erf}(x)=\frac{2}{\sqrt{\pi}}\int_{0}^{x}e^{-t^{2}}{\rm
d}t$ and $\int^{\infty}_{1/\tau^{2}}\frac{e^{-t}}{t}{\rm
   d}t\simeq-\gamma_{{\rm E}}-\alpha(1)n$ in the large $n$ limit, where
   $\gamma_{\rm E}=0.5772$ is the Euler constant. The theoretical predictions are
   compared with the numerical results in Fig.~\ref{fig:cycle}. The
   exponential growth of the typical cycle length is verified, which
   suggests that chaotic attractors exist in the state space of a
   randomly connected neural network.

\begin{figure}[h!]
\begin{tabular}{ll}
{\bf (a)} & {\bf (b)}\\
\includegraphics[bb=90 35 738
537,width=8cm]{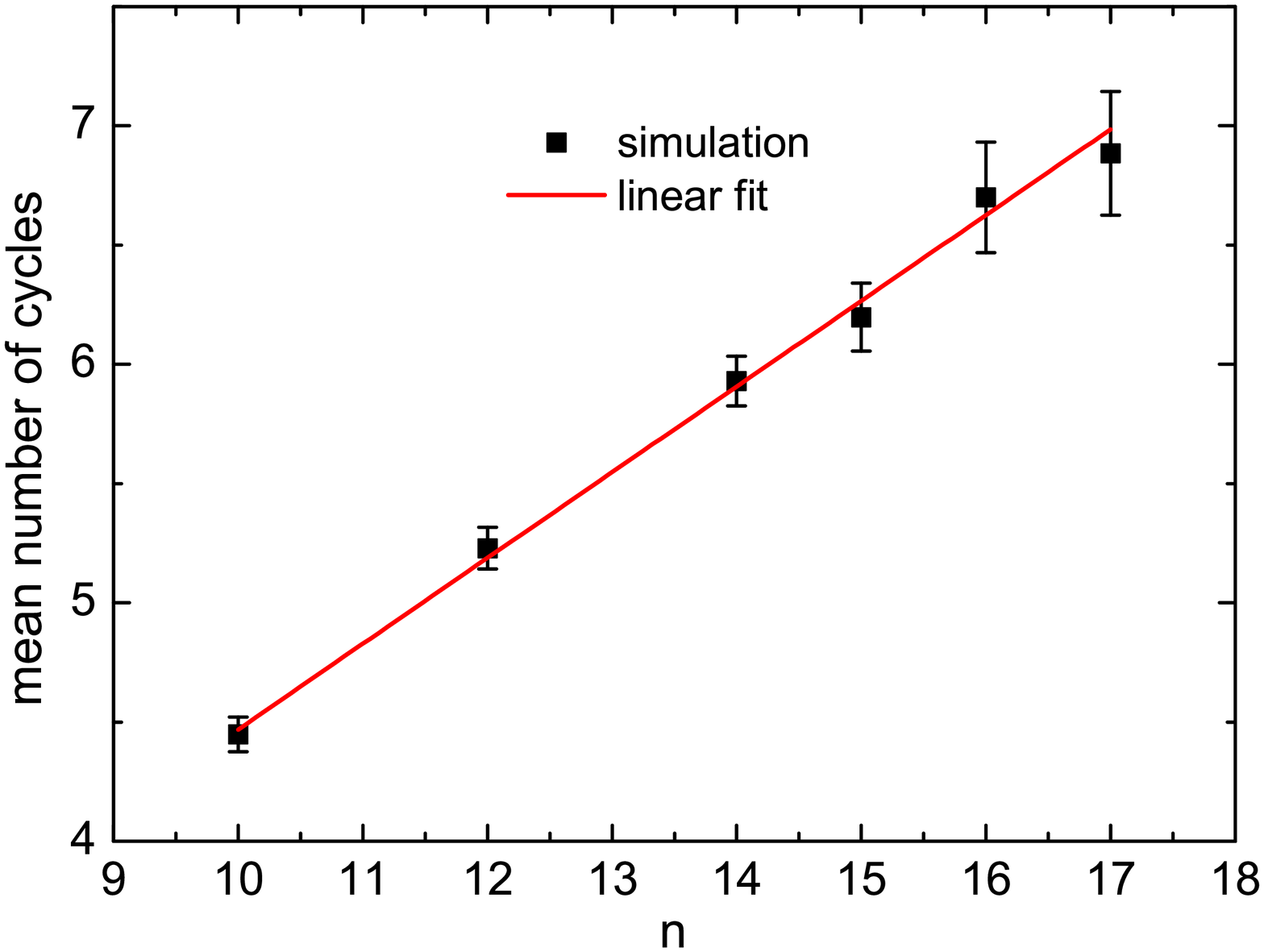}&
 \includegraphics[bb=67 20 740 545,width=8cm]{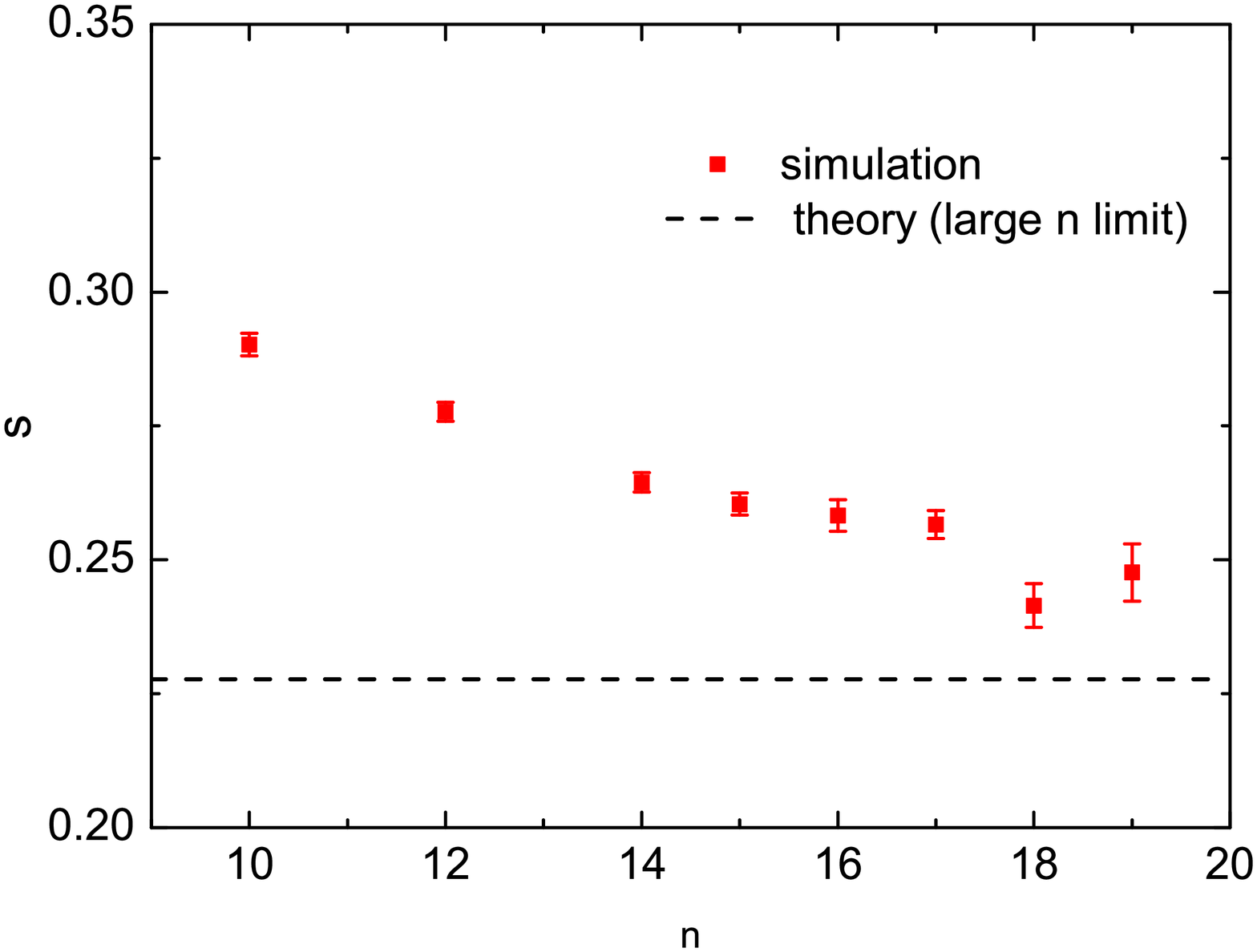}
 \end{tabular} \caption{(Color online) (a)
Linear dependence of the number of cycles on network size $n$. (b)
Entropy density of the attractive states defined by
$s=\frac{1}{n}\ln N_{{\rm att}}$. As $n$ increases, the numerical
data gets to the theoretical prediction.
  }\label{fig:entropy}
\end{figure}

Following the same spirit, one can derive the number of attractors
as $2^n Z \simeq-3\alpha(1) n/4-3\gamma_{\rm E}/4$, from which a
linear dependence~\cite{Young-1988,Berg-1992,Bastolla-1997} is
confirmed (see also Fig.~\ref{fig:entropy} (a), the linear fit gives
the slope $0.360\pm0.010$, compatible with the theoretical value
$0.342$). Another interesting quantity is the number of the
attractive states $N_{{\rm att}}$ belonging to all cycles (e.g., a
cycle of length $l$ has $l$ attractive states), which is expected to
grow exponentially with the network size $n$. This quantity is
evaluated by our theory as $N_{{\rm
att}}=2^{n}\sum_{l=1}^{2^n}P(l)$, and can be quantified as the
growth rate (entropy density)
$s=\lim_{n\rightarrow\infty}\frac{1}{n}\ln N_{{\rm att}}$.  In the
large $n$ limit, we obtain $s=-\alpha(1)/2$, which is compared with
the numerical results at finite $n$. As shown in
Fig.~\ref{fig:entropy} (b), as $n$ increases, $s$ decreases,
approaching the asymptotic limit $0.2277$.

The deviation at small $n$ (or at $l$ far from the characteristic
length, see Fig.~\ref{fig:cum}) comes from three approximations. One
is Eq.~(\ref{eq:pinit}), which becomes invalid at small $l$ where
$p_{\rm init}$ also depends on $l$, but Eq.~(\ref{eq:pinit}) becomes
reasonable for large $l$ (as occurs in our case where the typical
cycles are long). The second one is Eq.~(\ref{eq:cyc_length2}). This
approximation is valid in the range of $l$ specified by
Eq.~(\ref{eq:cond}). Note that this condition is consistent with the
numerical results shown in Fig.~\ref{fig:cum}. The last approximation is
Eq.~(\ref{eq:markov}), which breaks down for small $n$ at which two
or more time-steps memory should be considered. In the large network
size limit, these approximations become exact and the dynamics can
be described by a Markovian process in terms of the state overlap.
Thus, as we focus on the structure of attractors at a relatively
large but finite $n$, these effects are not significant.

\section{Conclusion}
\label{Conc} In this work, we studied the deterministic dynamics of
a randomly connected neural network and proposed a simple Markovian
stochastic process to describe the evolution of the overlap of two
states along the dynamics trajectories. The properties of the state
concentration can be studied by a mean field computation, and
furthermore, the theoretical cumulative distribution of cycle length
is compared with the numerical simulation results. The typical
length of cycles is predicted and observed to grow exponentially
with the network size. The number of attractive states on all cycles
has also an exponential growth with the network size, and its
typical value can also be predicted by our theory.

Our theory should have potential to be
generalized to treat more complex situations, e.g., couplings
between neurons are correlated, where one time-step memory is not
enough to describe the dynamics and strong memory effects induced by
retarded self-interaction could be incorporated by introducing a
back-action field (two time-steps memory)~\cite{Huang-2014}. The
current analysis is also restricted to the parallel type of dynamics,
whereas, the sequential
(asynchronous) dynamics seems to be more natural, and our current method may apply to this type of dynamics, although the
computation will become more complicated. However, the
statistical properties of attractors would not change
qualitatively, as expected from numerical simulations~\cite{Young-1988}.

Our work is expected to provide insights towards understanding how
the neural network processes information and stores temporal
sequences~\cite{Abbott-2005}, which will be left for future study.

\section*{Acknowledgments}

We are grateful to Ugo Bastolla, Naoki Masuda, Hiroyasu Ando, and
Shun-ichi Amari for useful discussions. This work was supported by
RIKEN Brain Science Institute and the Brain Mapping by Integrated Neurotechnologies for Disease
Studies (Brain/MINDS) by the Ministry of Education, Culture, Sports,
Science and Technology of Japan (MEXT).

\appendix
\section{Dynamic functional integral method}
\label{DynF}We compute the dynamics of the state overlap using the
dynamic functional integral (mean-field) method (see
\cite{Taro-2011} for similar calculations).  In this section, we
express time indices as lower case characters, \eg, $h_i(t)=h\ti$,
and follow the convention that summations are neglected if the same
indices appear twice in an expression, \eg,
$\sum_{j}J_{ij}\sigma_{jt}=J_{ij}\sigma_{jt}$.

Let us first
define the ensemble of state trajectories,
$\bh=\{h\ti|i=1,\dots,n,\, t=0,1,\dots\}$, averaged over different
networks:
\begin{eqnarray}
  \label{eq:Ph}
  P(\bh) \equiv \left[\prod_{i,t} \delta\left(h\ti-J_{ij}\sigma_{jt}\right)\right]_J,
\end{eqnarray}
where $\delta(\cdot)$ is the Dirac delta function and $[\cdot]_J$ is
the average over the random couplings. In the following, we denote by $\<\cdot\>$ an average
with respect to $P(\bh)$.

The joint
distribution of the overlap $\bq=\{q\ts\}_{t>s}$ is
\begin{eqnarray}
  \label{eqa:P1}
  P(\bq)&\equiv&\left\< \prod_{t> s}\delta\left(q\ts-\frac{1}{n}\sigma\tj\sigma\sj\right) \right\>\\
  &=&\int \left(\prod_{i,t} dh\ti\right) \left[\prod_{i,t}\delta(h\ti-J_{ij}\sigma_{jt})\right]_J \prod_{t> s}\delta\left(q\ts-\frac{1}{n}\sigma\tj\sigma\sj\right)\nn\\
  &=&\int \left(\prod_{i,t}\frac{dh\ti d\hh\ti}{2\pi}\right) \left[\exp\left(\ii\hh\ti h\ti-\ii\hh\ti J_{ij}\sigma_{jt}\right)\right]_J \prod_{t> s}\delta\left(q\ts-\frac{1}{n}\sigma\tj\sigma\sj\right)\nn\\
  &=&\int \left(\prod_{i,t}\frac{dh\ti d\hh\ti}{2\pi}\right) \exp\left(\ii\hh\ti h\ti-\frac{1}{2n}\hh\ti\hh\si\sigma_{jt}\sigma_{js}\right)\prod_{t> s}\delta\left(q\ts-\frac{1}{n}\sigma\tj\sigma\sj\right)\nn\\
  &=&\int \left(\prod_{i,t}\frac{dh\ti d\hh\ti}{2\pi}\right) \exp\left(\ii\hh\ti h\ti-\frac{1}{2}q\ts\hh\ti\hh\si\right)\prod_{t> s}\delta\left(q\ts-\frac{1}{n}\sigma\tj\sigma\sj\right)\nn\\
  &=& \int \left(\prod_{t>s}\frac{n d\hq\ts}{2\pi}\right)\int \left(\prod_{i,t}\frac{dh\ti d\hh\ti}{2\pi}\right) \exp\left(n\ii\hq\ts q\ts -\ii\hq\ts\sigma\tj\sigma\sj + \ii\hh\ti h\ti-\frac{1}{2}q\ts\hh\ti\hh\si\right)\nn\\
  &=& \int \left(\prod_{t>s}\frac{n d\hq\ts}{2\pi}\right) \exp(-n f(\bq,\hat\bq)),\nn
\end{eqnarray}
where we have defined the action
\begin{eqnarray}
  \label{eqa:f}
f(\bq,\hat\bq) \equiv  - \ii\hq\ts q\ts - \ln \int dH \exp\mathcal{L},\\
\mathcal{L} \equiv \ii \hh_t h_t
-\frac{1}{2}q\ts\hh_t\hh_s-\ii\hq\ts\sigma_t\sigma_s,\nn
\end{eqnarray}
and $dH\equiv \prod_t (dh_t d\hh_t)/(2\pi)$. In the derivation, we
have used the Fourier transformation of the delta function
$\delta(x)=\int d\hat{x}/(2\pi) \exp(\ii \hat{x} x)$ for each
delta function in Eq.~(\ref{eqa:P1}), and we have taken the average
over the independent Gaussian variables $\{J_{ij}\}$ of mean 0 and
variance $1/n$. For large $n$, the distribution of
Eq.~(\ref{eqa:P1}) is well approximated by a Gaussian distribution,
where the peak is specified by the saddle-point equations:
\begin{eqnarray}
  \label{eqa:spe}
  0&=&\frac{\@f}{\@q\ts}=-\ii \hq\ts+\frac{1}{2}\<\hh_t\hh_s\>_{\mathcal L},\\
  0&=&\frac{\@f}{\@\ii\hq\ts}=-q\ts+\<\sigma_t\sigma_s\>_{\mathcal L},\nn
\end{eqnarray}
with average $\<\cdot\>_{\mathcal L}\equiv(\int \cdot e^{\mathcal
  L}dH)/(\int e^{\mathcal L}dH)$.  We can easily see that $\hat\bq=0$
is a solution of Eq.~(\ref{eqa:spe}) \cite{Taro-2011}. Hence, if
$\hat\bq=0$, the average $\<\cdot\>_{\mL0}\equiv\<\cdot\>_{\mathcal
  L}|_{\hat \bq=0}$ is an average over Gaussian $\bh$ of mean
$\<h_t\>_{\mL0}=0$ and covariance $\<h_th_s\>_{\mL0}=q\ts$, which
simplifies the saddle-point equation of Eq.~(\ref{eqa:spe}) in terms
of a closed-form expression of $\bq$ by
\begin{eqnarray}\label{eqa:spe_q}
  q_{t+1,s+1}&=&\<\sigma_{t+1}\sigma_{s+1}\>_{\mL0}\\
  &=& \iint Dx Dy \,\sgn(x)\sgn\left(q\ts x+\sqrt{1-q\ts^2}y\right)\nn\\
  &=& \vp(q\ts),\nn
\end{eqnarray}
with $\vp(q)\equiv (2/\pi)\arcsin(q)$ and a Gaussian measure $Dx
\equiv \exp(-x^2/2)/\sqrt{2\pi}$.  Note that $|\vp(q)|\le |q|$ and the
equality holds only at $q=0$ and $q=\pm 1$. Hence, unless $q=1$
initially, the overlap rapidly converges in a few steps to zero
in the $n\to\infty$ limit.

 The
order parameters fluctuate around the saddle-point solution of
Eq.~(\ref{eqa:spe_q}) for finite $n$. This fluctuation of $\bq$ and
$\hat\bq$ is characterized to the leading order by the Hessian
matrix of $f$, \ie,
\begin{eqnarray}
  \label{eqa:hessian}
  A_{ts,t's'}&\equiv&\frac{\@^2f}{\@q\ts\@q_{t's'}}=\frac{1}{2}\frac{\@\<\hh_{t'}\hh_{s'}\>_{\mL0}}{\@q\ts},\\
  B_{ts,t's'}&\equiv&\frac{\@^2f}{\@q\ts\@\hq_{t's'}}=-\ii\delta_{t,t'}\delta_{s,s'}+\ii\frac{\@\<\sigma_{t'}\sigma_{s'}\>_{\mL0}}{\@q\ts},\nn\\
  C_{ts,t's'}&\equiv&\frac{\@^2f}{\@\hq\ts\@\hq_{t's'}}=\<\sigma_t\sigma_s\sigma_{t'}\sigma_{s'}\>_{\mL0}-\<\sigma_t\sigma_s\>_{\mL0}\<\sigma_{t'}\sigma_{s'}\>_{\mL0},\nn
\end{eqnarray}
for $t>s$ and $t'>s'$, where the Hessian matrix is evaluated at the saddle-point solution
of the order parameters, \ie, $\hat\bq=0$ and the solution of
Eq.~(\ref{eqa:spe_q}).

In the current setup, the Hessian matrix is simply given by
\begin{eqnarray}
  \label{eqa:hessian2}
  A_{ts,t's'}&=&0,\\
  B_{ts,t's'}&=&-\ii\delta_{t,t'}\delta_{s,s'}+\ii\vp'(q_{ts}) \delta_{t',t+1}\delta_{s',s+1},\nn\\
  C_{ts,t's'}&=&\delta_{t,t'}\delta_{s,s'}+O(q^2),\nn
\end{eqnarray}
for $t> s$ and $t'> s'$, where $\vp'(q)\equiv d\vp(q)/dq$.
Note that the $O(q^2)$ contribution in $C_{ts,t's'}$ can be more explicitly estimated, for example by applying Plackett's approximation~\cite{Bacon-1963}.
Here, we would like to evaluate the ($n$ multiplied) covariance of the overlap
parameter, $\tilde{A}_{ts,t's'}\equiv n \mbox{Cov}[q\ts,q_{t's'}]$.
By applying the matrix inversion lemma, we find that its inverse is
\begin{eqnarray}
  \label{eqa:hessian3}
  \tilde{A}^{-1} = A-B C^{-1} B^T = (\ii B)(\ii B)^T + O(q^2).
\end{eqnarray}
This relation indicates that for small $\bq$ the linear combination,
\begin{eqnarray}
  \label{eqa:eta}
  {\bm\eta} &\equiv& \sqrt{n}(\ii B)^T {\bm{\delta q}},
\end{eqnarray}
of the fluctuation of the overlap parameter, ${\bm{\delta q}}$, is
white Gaussian random variables. To see this, one can apply the
transformation of variables and find that
\begin{eqnarray}
  \label{eqa:eta2}
  P({\bm \eta}) &=& \int \delta\left({\bm\eta}- \sqrt{n}(\ii B)^T {\bm{\delta q}}\right) \exp\left(-\frac{n}{2}{\bm{\delta q}}^T \tilde{A}^{-1} {\bm{\delta q}}\right) d{\bm{\delta\bq}}\nn\\
&\sim& \exp\left(-\frac{1}{2}{\bm \eta}^T {\bm \eta}\right).
\end{eqnarray}
Thus, Eq.~(\ref{eqa:eta}) indicates that the finite-size fluctuations
 of the order parameter are described by
\begin{eqnarray}
\label{eqa:dq0}
\frac{1}{\sqrt{n}}\eta_{ts}&=&\left[\delta_{t',t}\delta_{s',s}-\vp'(q_{t's'})\delta_{t,t'+1}\delta_{s,s'+1}\right]\delta q_{t's'}\nn\\
&=&\delta q_{ts}-\vp'(q_{t-1,s-1})\delta q_{t-1,s-1}.
\end{eqnarray}
Altogether, summarizing that $q_{ts}=\vp(q_{t-1,s-1})$ in the $n\to\infty$ limit and that the finite-size correction is described by Eq.~(\ref{eqa:dq0}), we obtained, for finite $n$,
\begin{eqnarray}
  \label{eqa:dq}
q_{ts} = \vp(q_{t-1,s-1})+\frac{1}{\sqrt{n}}\eta_{ts} + O(q^2),
\end{eqnarray}
which is a simple Markovian process that involves white Gaussian noise of variance $1/n$.

Recalling the definition of the overlap parameter,
$q_{t+1,s+1}=\sgn(h_{it})\sgn(h_{is})/n$, and
that $h_i$ with different $i$ tend to become
 independent in the $n\to\infty$ limit, we know that the
overlap parameter must be distributed approximately according to a binomial
distribution. Extrapolating this observation, the result of
Eq.~(\ref{eqa:dq}) is consistent with the Markovian dynamics of
\begin{eqnarray}
  \label{eqa:markov}
  P(q_{t+1,s+1}) = \int W(q_{t+1,s+1}|q_{ts})P(q_{ts})dq_{ts}
\end{eqnarray}
with the binomial transition probability
\begin{eqnarray}
\label{eqa:transition} W\left(\left.
q_{t+1,s+1}=\frac{2m-n}{n}\right|q_{ts}\right) =
\binom{n}{m}\left(\frac{1+\vp(q_{ts})}{2}\right)^m
\left(\frac{1-\vp(q_{ts})}{2}\right)^{n-m},
\end{eqnarray}
where $m$ indicates the number of units taking the same state at time
$t+1$ and $s+1$.

In summary, this result shows that the Markovian dynamics of
Eq.~(\ref{eqa:markov}) provides a good approximation of the dynamics
of the overlap parameter once $O(q^2)$ terms become negligible near the stationary
state.
\section{The dynamics of the state overlap does not depend on the initial state $\sigma(0)$ }
\label{INIT} A specific choice of the initial state $\bs(0)$ is not
important to study dynamics of the state overlap for random ensemble
of networks as long as $\bs(0)$ is selected independently of the
network connections $\{J_{ij}\}$. Without losing generality, we can
set $\sigma_i(0)=1$ for all $i$.

To see this point, we consider a simple transformation of variables,
\begin{eqnarray}
  \tilde\sigma_i(t)=\sigma_i(t)\sigma_i(0).
\end{eqnarray}
The state overlap is also described in terms of these transformed
variables by $q\ts = (1/n) \sum_i
\tilde{\sigma}_i(t)\tilde{\sigma}_i(s)$ and the initial state is
given by $\tilde{\sigma}_i(0)=1$ for all $i$.

These transformed variables follow the same update rule as the
original one,
\begin{eqnarray}
  \tilde{\sigma}_i(t+1)=\sgn\left(\sum_j \tilde{J}_{ij}\tilde{\sigma}_j(t)\right),
\end{eqnarray}
except that the coupling matrix is given by
$\tilde{J}_{ij}=\sigma_i(0)J_{ij}\sigma_j(0)$ instead of $J_{ij}$.
Notably, the distribution of $\{\tilde{J}_{ij}\}$ is the same as
that of $\{J_{ij}\}$ as long as $\bs(0)$ is chosen independently of
$\{J_{ij}\}$. Therefore, to study the dynamics of the state overlap,
we can alternatively study the dynamics of these transformed
variables with the initial condition $\{\tilde\sigma_i(0)=1|
i=1,2,\dots,n\}$.

\section{The dynamics of the state overlap in random Boolean networks}
\label{RBN} The dynamics of the state overlap in random Boolean
networks is described by Eq.~(\ref{eq:dalpha}) with
$\vp_{BN}(q)=\delta_{q,1}$, which is simply
\begin{eqnarray}
  \label{eqa:dalpha_bn}
  \alpha_{t+1,s+1}(q)&=&\left\{\begin{array}{ll}
H(q)-\ln 2+\alpha\ts^*, & (q\ne 1)\\
\max\left\{\alpha\ts(1), -\ln 2+\alpha\ts^*\right\}, & (q=1)
\end{array}\right.
\end{eqnarray}
where $\alpha\ts^*\equiv \max_{q'\ne 1}\alpha\ts(q')$.
Let us assume that there is no perfect overlap of
states initially, \ie, $\mbox{Prob}(q_{l,0}=1)=0$. This means that
 $\alpha_{l,0}(1)<-\ln 2 + \alpha_{l,0}^*$ and
$\lim_{n\to\infty}\alpha_{l,0}^*=0$, because the initial overlap distribution
 $P_{l,0}(q)=\exp(n\alpha_{l,0}(q))$ must be normalized.
Thus, the
dynamics of Eq.~(\ref{eqa:dalpha_bn}) converges in one step to a
stationary solution
\begin{eqnarray}
  \label{eqa:dalpha_sol_bn}
  \alpha(q)=H(q)-\ln 2.
\end{eqnarray}
Moreover, we have from Eq.~(\ref{eq:beta})
\begin{eqnarray}
  \label{eqa:beta_bn}
  \beta(q')&=&\ln \frac{1+\vp_{BN}(q')}{2}+H(q')\\
&=&\left\{\begin{array}{ll}
H(q')-\ln 2, & (q'\ne 1)\\
0. & (q'= 1)
\end{array}\right.\nn
\end{eqnarray}
This indicates that states mainly concentrate from $q=0$ if they do not already concentrate.

This analysis also provides important
information about the eigenvalues of the transition matrix $W$ at
the large network size limit. The first eigenvalue is trivial, $\lambda_1=1$,
with the eigenfunction $f_1(q)=\delta_{q,1}$, indicating that states never separate
once they concentrate. The second eigenvalue, $\lambda_2\approx 1$, is a non-trivial one that
corresponds to the quasi-stationary state with the eigenfunction $f_2(q)=\exp(n\alpha(q))$,
where $\alpha(q)$ is given by Eq.~(\ref{eqa:dalpha_sol_bn}). The other eigenvalues $\lambda_k$ for $k\ge 3$
are all zero because the distribution of the overlap converges in a single step to the quasi-stationary state.
Furthermore, the fact that state concentration happens with probability $2^{-n}$ at each time step
suggests that $\lambda_2=1-2^{-n}$.

\section{Simulation details of the dynamics}
\label{NumS} The total number of states is $2^{n}$. They form a
state set called $\mathcal {S}$. We also denote a path set $\mathcal
{P}$ recording the states on a dynamics trajectory. Only the state
index is stored in both sets.
\begin{description}
  \item[Step 1.] Choose the first state $\boldsymbol{\sigma}^{1}$ in $\mathcal {S}$ as a starting point for the parallel dynamics, and remove this state from $\mathcal {S}$ at
  the same time.
  \item[Step 2.] $\boldsymbol{\sigma}^{1}$ evolves to $\boldsymbol{\sigma}'$ by one step of the parallel dynamics (all neurons' states are updated for one time).
  \begin{description}
    \item[Step 2.1.] If $\boldsymbol{\sigma}'\in\mathcal {S}$,
    remove it from $\mathcal {S}$, put the index of $\boldsymbol{\sigma}'$ into $\mathcal {P}$, and continue to perform the parallel
    dynamics, i.e., let $\boldsymbol{\sigma}^{1}=\boldsymbol{\sigma}'$, then go to \textbf{Step 2};
    \item[Step 2.2] Otherwise, compare $\boldsymbol{\sigma}'$ with
    the one in the set $\mathcal {P}$ and if they coincide with each other, a new
    cycle is identified and the length is recorded at the same time, then go to  \textbf{Step 3}; otherwise, no
    new cycle is found and go to \textbf{Step 3}.
  \end{description}
\item[Step 3.] Go to \textbf{Step 1} until the set $\mathcal {S}$
becomes empty.
\end{description}


\end{document}